\documentclass[conference]{IEEEtran}
\IEEEoverridecommandlockouts
\usepackage{cite}
\usepackage{amsmath,amssymb,amsfonts}
\usepackage{algorithmic}
\usepackage{graphicx}
\usepackage{textcomp}
\usepackage{xcolor}
\usepackage{amsmath,amssymb,amsthm,mathtools}
\usepackage{booktabs,tabularx,array}
\usepackage{booktabs,longtable,array,amsmath}
\usepackage{url}
\usepackage[numbers]{natbib}
\usepackage{pgfplots}
\usepackage{enumitem}
\usepackage{microtype}
\usepackage{fancyhdr}
\setlist{nosep,leftmargin=*}

\def\BibTeX{{\rm B\kern-.05em{\sc i\kern-.025em b}\kern-.08em
    T\kern-.1667em\lower.7ex\hbox{E}\kern-.125emX}}

\newcommand \modelname{TraGR}
\begin{document}

\title{From Tokens to Behaviors: Rethinking Generative Recommendation for User Trajectory Prediction}

\author{
\IEEEauthorblockN{
Jiaqi Wang$^{1,2,*}$,
Hanchen Yang$^{3,*}$,
Hengyu Zhang$^{2}$,
Hongxu Ma$^{3}$,
Wenqi Rao$^{3}$, 
Hongyang Wang$^{3}$, \\
Zhongbo Sun$^{3}$,  
Renliang Qian$^{3}$, 
Weixin Chen$^{3}$,
Wengen Li$^{1}$, 
Jihong Guan$^{1}$,  \\
Shuigeng Zhou$^{4}$,
Qingpeng Cai$^{3,\dagger}$,
Peng Jiang$^{3}$,
Hong Cheng$^{2,\dagger}$
}
\IEEEauthorblockA{
$^{1}$Tongji University, Shanghai, China
\quad
$^{2}$The Chinese University of Hong Kong, Hong Kong SAR, China\\
$^{3}$Kuaishou Technology, Beijing, China
\quad
$^{4}$Fudan University, Shanghai, China\\
\texttt{caiqingpeng@kuaishou.com},
\texttt{hcheng@se.cuhk.edu.hk}
}
\thanks{
$^{*}$Jiaqi Wang and Hanchen Yang contributed equally to this work.
$^{\dagger}$Corresponding authors: Hong Cheng and Qingpeng Cai.
}
}
\maketitle

\begin{abstract}

User behavior trajectory prediction (UBTP) aims to capture how a user’s needs evolve over time by predicting the sequence of future interactions from that user’s interaction history.
Recent generative recommendation~(GR) methods address this task through autoregressive generation of semantic IDs (SIDs).  
However, serial SID decoding couples token generation with behavior state transitions and requires multiple dependent steps per interaction. Discrete item feedback overlooks predictive uncertainty, and teacher-forced training leaves a training–inference gap in trajectory generation.
To address these limitations, we propose \modelname{}, a behavior-aware generative recommendation framework that aligns generation, feedback, and learning with successive user behaviors. Its behavior-level architecture predicts SID tokens in parallel within each interaction and models behavioral transitions through recurrent state updates, with joint SID and item-level supervision. Probability feedback combines the selected behavior with a probability-weighted summary of alternatives to guide subsequent predictions. Relation-aware trajectory learning incorporates collaborative transition signals and later credit assignment into training on model-generated trajectories. 
Experiments on four public datasets demonstrate that \modelname{} consistently outperforms state-of-the-art baselines across different prediction horizons. Code: https://github.com/CalmaQi/GR4UBT.
\end{abstract}


\section{Introduction}
\label{sec:introduction}

\begin{figure}[t]
    \centering
    \includegraphics[width=\linewidth]{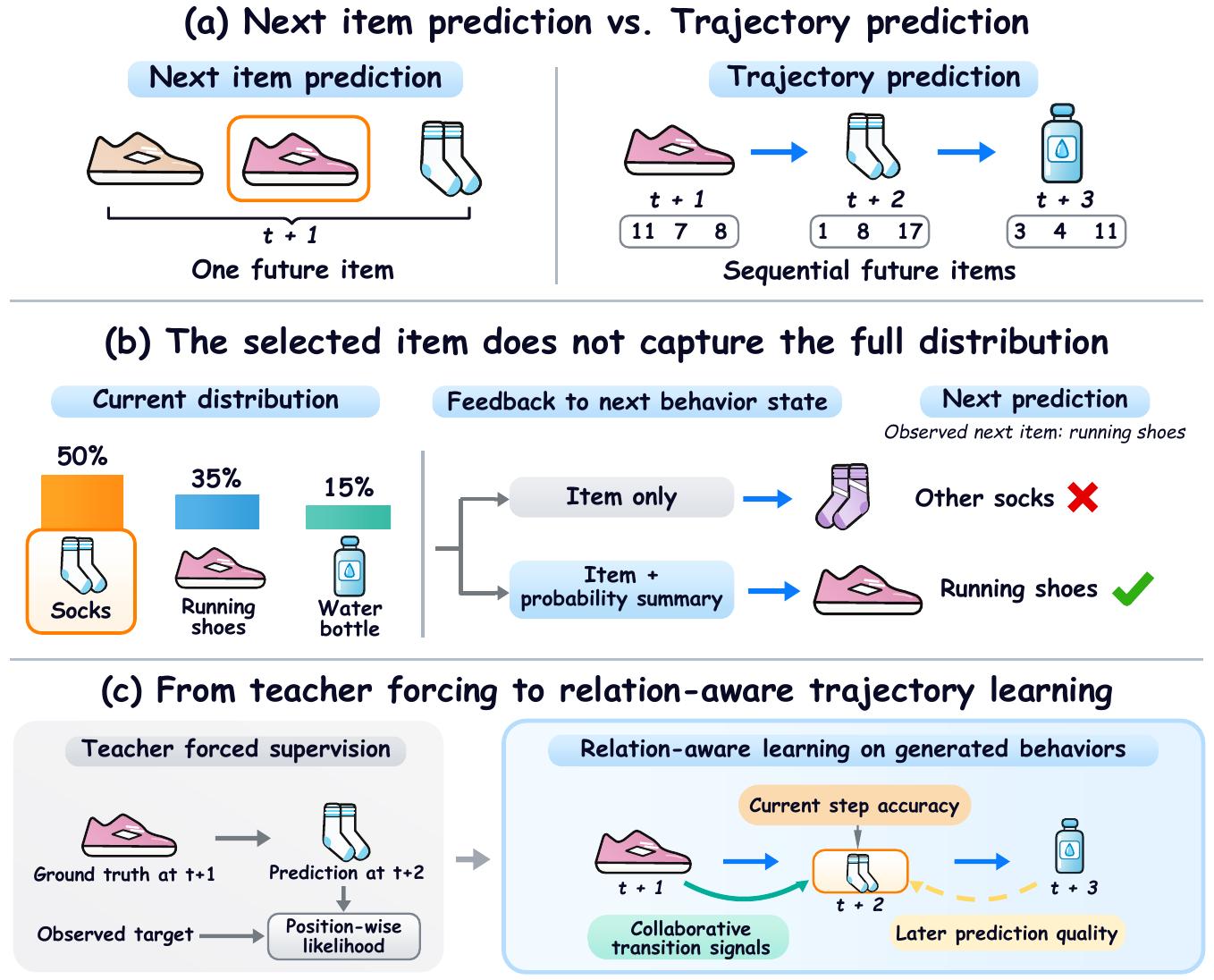}
    \caption{Generative recommendation for user trajectory prediction.  (a) Next item candidates are alternatives for one interaction, whereas a trajectory contains items at successive future positions.  (b) With socks selected, item-only feedback may predict more socks, while a probability summary retains support for running shoes and guides the next prediction.  (c) Model-generated trajectories require relation-aware learning to address the training–inference gap.}
    \label{fig:motivation}
\end{figure}

Recommender systems play an important role in helping users discover relevant items from large catalogs~\cite{liu2024sequential,11629195,liu2026community,zhang2024id}. Personalized recommendation mainly needs to learn what users like from their past interactions. Since these interactions unfold over time, sequential recommendation has been developed to learn from historical user–item interactions and predict the next item a user is likely to interact with, yielding remarkable success in both academia and industry~\cite{ning2024user,huang2019online,pancha2022pinnerformer,peng2022mixed,li2025personalized}. Recent advances in large language models (LLMs)~\cite{achiam2023gpt,guo2025deepseek,yang2025darec} have popularized generative recommendation (GR), shifting recommendation from scoring items to generating their identifiers. A prominent approach represents items as Semantic IDs (SIDs) and autoregressively generates their constituent tokens conditioned on the user's interaction history~\cite{rajput2023recommender}. This formulation brings user preference modeling and item prediction into a unified sequence generation framework.

Despite these advances, most sequential and generative recommendation methods remain focused on next item prediction~\cite{IJCAISurvey,ChenLLMX22,XieSLWGZDC22,hao2024meta}. While effective for anticipating an immediate interaction, this objective provides only a limited view of how user behavior unfolds over time. Understanding and predicting a user's subsequent course of actions is important for anticipating related needs and enabling more proactive recommendation. For example, a user may purchase running shoes, followed by sports socks and a water bottle, as illustrated in Fig.~\ref{fig:motivation}(a). Predicting only the next purchase captures one part of this progression, whereas forecasting the sequence reveals how related behaviors follow one another. User Behavior Trajectory Prediction (UBTP) aims to capture this broader progression by forecasting an ordered sequence of future interactions from a user's history~\cite{huang2025listwise,huang2026generative}. Unlike a next item ranked list, whose items are alternative candidates for a single future interaction, a predicted trajectory contains behaviors at successive future positions.

Recent work has begun to explore this direction. LPDO~\cite{huang2025listwise} employs diffusion with listwise preference optimization to model future trajectories. MaskGR~\cite{shah2025masked} explores multi-item recommendation through masked diffusion over SIDs. EchoRec~\cite{ma2026echorec} combines supervision from future interactions with cycle-consistent preference alignment to enhance generative recommendation. GCB~\cite{huang2026generative} extends SID-based GR to trajectory prediction, autoregressively generating the semantic tokens of successive items so that earlier generated behaviors provide context for later predictions. However, extending SID-based generation from a single item to a behavior trajectory also exposes limitations in decoding, information propagation, and learning. Token-autoregressive decoding intertwines within-item SID generation with across-item behavior transitions. Passing selected items forward without explicit predictive uncertainty feedback leaves useful information about alternative behaviors behind. Moreover, teacher-forced likelihood training learns to continue observed behavior prefixes, whereas inference requires the model to continue its own predictions. We identify the following three key limitations.

\textbf{First, a mismatch between token-level generation and behavior-level dependencies.}  Serial decoding uses the same autoregressive process to construct each item's semantic code and move between successive items~\cite{luo2026awakening}. This requires multiple dependent decoding steps per behavior and leaves transitions between complete interactions implicit in the token sequence.
\textbf{Second, the neglect of predictive uncertainty during trajectory rollout.} Autoregressive methods typically pass selected items to later predictions~\cite{huang2026generative}, although these intermediate behaviors have not yet been observed. In Fig.~\ref{fig:motivation}(b), socks are selected with probability 50\%, while running shoes also receive 35\%. The socks embedding alone does not convey this alternative support: it may lead to another sock prediction, whereas the candidate probabilities provide useful context for predicting running shoes next.
\textbf{Third, lack of relation-aware learning to bridge the training–inference gap in trajectory generation.} Models learn from observed prefixes in training but rely on their own predictions at inference~\cite{feng2026psg,cao2026kuaishou,huang2026generative}. In Fig.~\ref{fig:motivation}(c), the behavior at $t+2$ should be evaluated by its current accuracy, the collaborative transition from $t+1$, and its contribution to predicting $t+3$. Yet likelihood supervision under observed prefixes does not directly evaluate these generated transitions or assign later prediction rewards to earlier choices.

To address these limitations, we propose \modelname{}, a behavior-aware generative recommendation framework that aligns generation, feedback, and learning with successive user behaviors.
To resolve the granularity mismatch, we build a \emph{behavior-level generation architecture} on OPQ tokenization and parallel SID prediction. At each step, parallel SID predictions define a distribution over items, and a recurrent behavioral state advances once per item interaction. Joint SID and item-level supervision learns semantic representations and ordered behavior dependencies.
To retain predictive uncertainty during trajectory generation, we introduce \emph{probability feedback}. This mechanism combines the selected item's representation with a candidate summary weighted by both predicted probabilities and similarity to the selected item. The resulting feedback embedding updates the behavioral state, allowing subsequent predictions to use both the selected behavior and information about its plausible alternatives.
To bridge the training–inference gap in trajectory generation, we introduce \emph{relation-aware trajectory learning}. After supervised initialization, the model generates trajectories conditioned on its own predictions. Current candidates are evaluated using observed future targets and directed collaborative transition signals, while later credit assigns subsequent prediction quality to earlier choices. These signals jointly optimize the SID predictor and behavior-state transitions, enabling behavior-level dependencies to be reflected throughout the generation process. 

Our main contributions are summarized as follows:
\begin{enumerate}
    \item \textbf{Behavior-level trajectory generation.}   We integrate parallel SID prediction with recurrent behavior state updates, trained under joint SID and item-level supervision.
    \item \textbf{Probability feedback.}  We pass the selected item and a probability-weighted summary of related candidates to the next step, allowing subsequent predictions to use information about current alternatives. 

    \item \textbf{Relation-aware trajectory learning.} We combine current candidate evaluation with later credit assignment on model-generated prefixes, learning how each prediction follows preceding behaviors and supports later forecasts.
    \item \textbf{Comprehensive experimental evaluation.} We conduct extensive experiments on four public datasets and demonstrate strong predictive performance against competitive baselines. 
\end{enumerate}

\section{Related Work}
\label{sec:related-work}

\subsection{Generative Recommendation}
\label{subsec:related-generative}
Generative recommendation produces item identifiers conditioned on user history. Semantic IDs represent an item with several discrete codes, allowing different items to share semantic components~\cite{hong2025eager}. TIGER~\cite{rajput2023recommender} constructs hierarchical SIDs through residual quantization and autoregressively generates the codes of the next item. Subsequent work improves how these identifiers incorporate recommendation signals. LC-Rec~\cite{zheng2024adapting}  aligns language and collaborative semantics through learned item indices and alignment tasks, while LETTER~\cite{letter} incorporates semantic, collaborative, and code-diversity regularization into item tokenization. RPG~\cite{hou2025generating} addresses the cost of sequential code generation by constructing SIDs with optimized product quantization and predicting their code positions in parallel.
Other studies introduce collaborative guidance, interaction timing, and future behavior supervision into SID-based recommendation. TCBSA~\cite{shan2026from} projects item scores from a discriminative recommender onto SID prefixes to guide beam selection at test time, keeping both models frozen. ChronoSID~\cite{huang2026beyond} introduces time-gap supervision into item representation learning and interleaves historical time-gap tokens with SIDs to account for elapsed time between interactions. EchoRec~\cite{ma2026echorec} connects prediction branches across future positions and uses cycle-consistent preference alignment to learn from multiple future interactions. GCB~\cite{huang2026generative} applies SID generation to predict behavior trajectories. Building on these developments, we study how SID-based generative recommendation can guide the transition between predicted behaviors and support learning from the  trajectories.

\subsection{User Behavior Trajectory Prediction}
\label{subsec:related-trajectory}

User behavior trajectory prediction forecasts several successive interactions conditioned on an observed history~\cite{wu2024multi}. Each output position corresponds to a future interaction, making both positional prediction accuracy and dependencies between successive behaviors relevant to the task. 
LPDO~\cite{huang2025listwise} develops a diffusion-based approach and incorporates listwise preference supervision into the prediction of future interaction sequences. MaskGR~\cite{shah2025masked} also explores multi-item recommendation through masked diffusion over SIDs. EchoRec~\cite{ma2026echorec} combines supervision from future interactions with cycle-consistent preference alignment to enhance generative recommendation. GCB~\cite{huang2026generative} models future interactions as a concatenated SID sequence and trains its autoregressive generator with teacher-forced token likelihood, using preceding item tokens as context for subsequent predictions.
\modelname{} develops a unified behavior-level formulation for trajectory generation, feedback, and learning. It separates parallel SID prediction within an item from state updates between interactions. Each update receives the selected item together with a probability-weighted summary of its alternatives, allowing the current prediction distribution to inform subsequent predictions. Beyond supervised learning from observed trajectories, we optimize trajectories generated from the model's own previous predictions.

\subsection{Multi-Item Recommendation}
\label{subsec:related-multi-item}
Multi-item recommendation includes several tasks with different output units. Next-basket methods predict the products associated with the next shopping event. ReCANet~\cite{ariannezhad2022recanet} explicitly models users' repeat-purchase patterns, while CASE~\cite{cao2026case} models item-specific repurchase cycles in calendar time and uses set attention to identify items likely to be purchased again. Although these methods use temporal information from historical purchases, their target is a set of items within the next basket, without a prescribed interaction order among those items.
Session-wise recommendation and list- or page-level reranking organize multiple items into a recommendation output. OneRec~\cite{deng2025onerec} generates session-wise video lists for a user request and refines them through preference alignment. APCR~\cite{luo2025anchor} uses LLM-based pairwise comparisons against a reference item and position-aware reranking to refine recommendation lists. PAR~\cite{xi2023bird} performs page-level reranking by considering interactions within and across lists, together with the spatial relationships among displayed items. These methods aim to improve the quality of a recommended list or page. 

UBTP defines its output as a sequence of future user interactions. Each output position predicts the item at a specific future interaction. These interactions need not form a single basket or served session. The prediction target therefore differs from basket membership or page placement: UBTP forecasts which items the user will interact with and in what temporal order. In \modelname{}, intermediate items are predictions made before the corresponding interactions are observed, and they provide context for subsequent forecasts. We accordingly learn how each prediction follows the preceding behavior and how it supports accurate predictions of later interactions.

\section{Method}
\label{sec:method}

\subsection{Task Definition and Overview}
\label{subsec:revised-overview}

Let $\mathcal I=\{1,\ldots,N\}$ denote the item catalog, $H=(i_1,\ldots,i_n)$ an observed interaction history, and $Y=(y_1,\ldots,y_K)$ the next $K$ interactions in order. Given $H$, \modelname{} predicts an ordered trajectory $X=(x_1,\ldots,x_K)$, where $x_t$ forecasts the item at the $t$-th future interaction.

\modelname{} maintains a behavioral state that advances once per predicted item. At each step, parallel SID heads produce a distribution over items. The selected item and the current step's predicted distribution then update the state for the next prediction. Stage I jointly learns SID and item prediction. Stage II learns from model-generated trajectories: it evaluates how each candidate follows the preceding behavior and assigns later prediction quality to earlier choices, with KL regularization to a frozen reference model. At inference, items are selected greedily while the full predicted distribution continues to inform the state update.

\begin{figure*}[t]
    \centering
    \includegraphics[width=\linewidth]{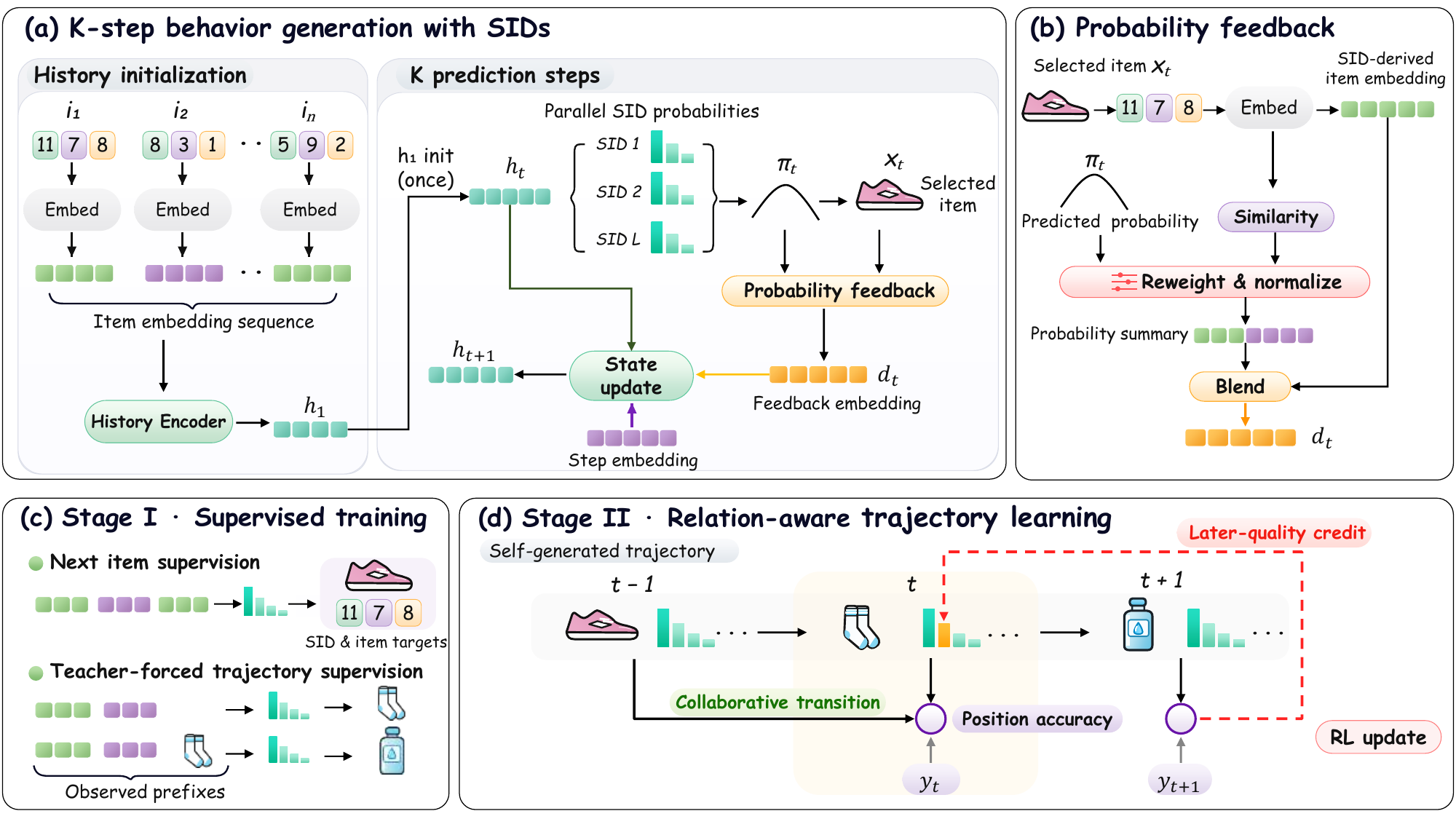}
    \caption{Overview of the proposed \modelname{}. (a) K-step behavior trajectory generation with parallel SID prediction and behavior-level state updates. (b) Probability feedback combines the selected item’s embedding with a feedback embedding weighted by predicted probabilities and similarity to the selected item. (c) Stage I jointly learns next item prediction and teacher-forced trajectory prediction. (d) Stage II learns from model-generated trajectories using position accuracy and collaborative transition rewards, later credit assignment, and KL regularization toward a frozen reference.}
    \label{fig:ubt4gr-framework}
    \label{fig:placeholder}
\end{figure*}

\subsection{Trajectory Generation with Semantic IDs}
\label{subsec:revised-generation}

\paragraph{SIDs and history encoding}
Following RPG~\cite{hou2025generating}, we apply optimized product quantization (OPQ) to item text features. Each item $i$ receives a SID $c(i)=(c_1(i),\ldots,c_L(i))$, with one token from each of $L$ codebooks. Let $E_{\ell,v}\in\mathbb R^d$ be the learned embedding of token $v$ at SID position $\ell$. The item representation is the mean of its token embeddings:
\begin{equation}
 e_\theta(i)=\frac{1}{L}\sum_{\ell=1}^{L}E_{\ell,c_\ell(i)}.
 \label{eq:revised-item-embedding}
\end{equation}
The SID assignments remain fixed while the embedding tables are learned. A causal Transformer encoder initializes the behavioral state from the history  sequence:
\begin{equation}
 h_1=\operatorname{Enc}_\theta
 \bigl(e_\theta(i_1),\ldots,e_\theta(i_n)\bigr)_n,
 \label{eq:revised-history}
\end{equation}
where the subscript $n$ selects the last history output and $\theta$ denotes the trainable generator parameters.

\paragraph{Parallel SID prediction}
We follow RPG's parallel prediction design and use a separate residual head for each SID position. All heads receive the same behavioral state $h_t$:
\begin{equation}
 z_{t,\ell}=h_t+f_\ell(h_t),
 \qquad \ell=1,\ldots,L,
 \label{eq:revised-sid-head}
\end{equation}
where $f_\ell$ is an MLP with output dimension $d$. The heads have separate parameters across SID positions and are shared  across behavior steps.
The SID-token probabilities are
\begin{equation}
 q_{t,\ell}(v)=
 \operatorname{softmax}_{v}\!
 \left( z_{t,\ell}^{\mathsf T} E_{\ell,v}/
 \tau_{\mathrm{sid}}\right),
 \label{eq:revised-code}
\end{equation}
where the softmax is over the vocabulary of codebook $\ell$. We combine the token probabilities assigned to each item's SID:
\begin{align}
 s_t(i)
 &=\frac{1}{L}\sum_{\ell=1}^{L}
 \log q_{t,\ell}(c_\ell(i)),
 \label{eq:revised-score}\\
 \pi_t(i)
 &=\frac{\exp(s_t(i)/\tau_{\mathrm{item}})}
 {\sum_{j\in\mathcal I}\exp(s_t(j)/\tau_{\mathrm{item}})}.
 \label{eq:revised-policy}
\end{align}
The positive temperatures $\tau_{\mathrm{sid}}$ and $\tau_{\mathrm{item}}$ control token and item normalization, respectively. All SID positions are predicted in parallel, and the resulting item distribution is normalized over the catalog. The catalog softmax cancels the codebook-specific log-normalization constants. Consequently, the item distribution is equivalent to a softmax over additive SID logits. 
\paragraph{Probability feedback}
\label{subsec:revised-feedback}
After selecting $x_t$, we combine its representation with a summary of plausible candidates. Let $b_i\in\mathbb R^d$ denote the representation used for feedback. Stage I uses the current mean-pooled SID representation $b_i=e_\theta(i)$. We weight each candidate by its predicted probability and its similarity to the selected item:
\begin{equation}
 w_t(i)=
 \frac{\pi_t(i)\exp\!
 \left( b_i^{\mathsf T} b_{x_t}/\tau_{\mathrm{sim}}\right)}
 {\sum_{j\in\mathcal I}\pi_t(j)\exp\!
 \left( b_j^{\mathsf T} b_{x_t}/\tau_{\mathrm{sim}}\right)}.
 \label{eq:revised-weights}
\end{equation}
Here $\tau_{\mathrm{sim}}>0$ controls the similarity weighting. The candidate probability summary and feedback vector are
\begin{equation}
 \mu_t=\sum_{i\in\mathcal I}w_t(i)b_i,
 \qquad
 d_t=(1-\eta)b_{x_t}+\eta\mu_t,
 \label{eq:revised-feedback}
\end{equation}
where $0\leq\eta<1$ controls the summary's contribution. The summary is computed over the full catalog, while the selected item retains an explicit contribution to $d_t$. When $\eta>0$, the dependence of the weights on $\pi_t$ allows later predictions to send gradients to earlier SID probabilities through this feedback.

\paragraph{State update}
A residual state transition head uses the feedback to advance the behavioral state once per item:
\begin{equation}
 h_{t+1}=h_t+\operatorname{MLP}_{\mathrm{tr}}
 \bigl([\operatorname{LN}(h_t);
        \operatorname{LN}(d_t);o_t]\bigr),
 \label{eq:revised-transition}
\end{equation}
where $\operatorname{LN}$ denotes layer normalization, $[\cdot;\cdot;\cdot]$ denotes concatenation, and $o_t$ is a learned embedding of future step $t$. The transition head is shared across steps and is applied for $t<K$. Thus, the SID heads operate in parallel within an item, while the transition head connects successive item predictions.

The resulting distribution $\pi_t$ depends on $H$ and the selected prefix $x_{<t}=(x_1,\ldots,x_{t-1})$ through the recurrent state. The trajectory distribution is
\begin{equation}
 P_\theta(X\mid H)=\prod_{t=1}^{K}
 \pi_\theta(x_t\mid H,x_{<t})
 =\prod_{t=1}^{K}\pi_t(x_t).
 \label{eq:revised-joint}
\end{equation}
Training with observed prefixes, training with sampled prefixes, and greedy inference use this same generation process with different item selection rules.

\subsection{Stage I: Joint Item and Trajectory Supervision}
\label{subsec:revised-supervised}

Stage I jointly optimizes next item prediction and teacher-forced trajectory prediction. Both objectives supervise the generator using items observed in the training sequences.

\paragraph{Next item supervision}
Let $(H^-,i^+)$ denote history--item pairs extracted from the training sequences. 
For each pair, the observed next item $i^+$ supplies two targets: its SID tokens $c_1(i^+),\ldots,c_L(i^+)$ supervise the parallel token distributions, and its item identity supervises the catalog distribution. Both predictions are computed from $H^-$. The loss is
\begin{equation}
\small{
 \mathcal L_{\mathrm{next}}
 =-\mathbb E
 \Bigg[
 \frac{1}{L}\sum_{\ell=1}^{L}
 \log q_{1,\ell}(c_\ell(i^+))
 +\lambda_{\mathrm{item}}\log\pi_1(i^+)
 \Bigg].}
 \label{eq:revised-next-loss}
\end{equation}
The token term trains each SID position against its codebook vocabulary. The item term trains the target item's probability over the catalog through the composed SID scores. Both terms update the same SID prediction heads. The coefficient $\lambda_{\mathrm{item}}$ weights item supervision.

\paragraph{Trajectory supervision}
Let $\mathcal D_{\mathrm{traj}}$ contain history--future pairs $(H,Y)$ with $K$ observed future interactions. At step $t$, the model predicts $\pi_t$ from $H$ and the preceding observed items $y_{<t}$. After evaluating the target probability $\pi_t(y_t)$, we use $y_t$ in the feedback and state transition for the next step. The trajectory loss is
\begin{equation}
 \mathcal L_{\mathrm{traj}}
 =-\frac{1}{K}
 \mathbb E_{(H,Y)\sim\mathcal D_{\mathrm{traj}}}
 \left[\sum_{t=1}^{K}\log\pi_t(y_t)\right].
 \label{eq:revised-traj-loss}
\end{equation}
Its item labels are the future interactions in the training sequence, and gradients propagate through the state transitions.

We jointly optimize the encoder, token embeddings, SID heads, and transition head using
\begin{equation}
 \mathcal L_{\mathrm I}
 =\mathcal L_{\mathrm{next}}
 +\lambda_{\mathrm{traj}}\mathcal L_{\mathrm{traj}},
 \label{eq:revised-stage-one}
\end{equation}
where $\lambda_{\mathrm{traj}}$ weights trajectory supervision. During Stage I, we set $\eta=0$, so each teacher-forced transition uses the observed item's current SID representation.

\paragraph{Initialization for Stage II }
After joint supervised training, we cache the feedback table $b_i$ from the learned SID embeddings. We initialize Stage II with the supervised generator, enable probability feedback with $\eta=0.75$, and freeze a copy of the generator with this configuration as the reference model $P_0$. The feedback is also used at inference.

\subsection{Stage II: Relation-Aware Trajectory Learning}
\label{subsec:revised-optimization}

Stage II optimizes the generator on trajectories sampled from its current policy. For each training pair $(H,Y)$, we generate $X\sim P_\theta(\cdot\mid H)$ autoregressively by sampling each $x_t$ from the catalog distribution $\pi_t$ and updating the behavioral state through probability feedback. The observed future $Y$ is used only for reward evaluation after generation. The objective considers both the compatibility of each prediction with the preceding behavior and its contribution to subsequent prediction quality.

\paragraph{Prediction and transition rewards}
For a candidate item $i$ at step $t$, we use
\begin{equation}
 r_t(i)=\frac{1}{K}
 \left[R(i,y_t)+\lambda_A A(x_{t-1},i)\right],
 \qquad x_0=i_n.
 \label{eq:revised-reward}
\end{equation}
Here $R(i,y_t)$ evaluates the prediction against the observed item $y_t$. The global collaborative transition score $A(i,j)$ measures support for the directed transition $i\rightarrow j$, using interactions aggregated across users in the training data.  The coefficient $\lambda_A \in [0,1]$ weights this transition reward. 

We define the position accuracy reward, $R(i,y_t)=\mathbb{I}[i=y_t]$, where $\mathbb{I}[\cdot]$ is the indicator function. Its expectation under the current policy is $\sum_{i\in\mathcal{I}}\pi_t(i)R(i,y_t)=\pi_t(y_t)$, so maximizing this term directly increases the probability assigned to the observed target.
To construct the collaborative transition score, we count ordered pairs of consecutive items in each user's chronologically ordered training sequence. Let $N_{ij}$ denote the number of times item $j$ immediately follows item $i$, aggregated across users, and let $N_i^{\mathrm{out}}=\sum_{v\in\mathcal{I}}N_{iv}$ be the total number of observed outgoing transitions from $i$. We define
\begin{equation}
A(i,j)=
\begin{cases}
N_{ij}/N_i^{\mathrm{out}}, & N_i^{\mathrm{out}}>0,\\
0, & N_i^{\mathrm{out}}=0.
\end{cases}
\label{eq:collaborative-transition-score}
\end{equation}
Thus, $A(i,j)$ is the empirical frequency of choosing $j$ as the next item after $i$, conditional on an observed outgoing transition. Unobserved transitions receive zero additional reward. The scores are computed using only the training data.

The two terms provide complementary signals: $R(i,y_t)$ keeps the prediction aligned with the observed target, while $A(x_{t-1},i)$ favors continuations supported by behavior patterns across users. The transition term is evaluated against the last item in the generated prefix, with the final historical item providing the predecessor at the first prediction step.

We maximize the expected total reward
\begin{equation}
 J(\theta)=
 \mathbb E_{(H,Y)\sim\mathcal D_{\mathrm{traj}}}
 \mathbb E_{X\sim P_\theta(\cdot\mid H)}
 \left[\sum_{t=1}^{K}r_t(x_t)\right].
 \label{eq:revised-utility-objective}
\end{equation}
The dependence of $r_t$ on the sampled prefix and observed future is left implicit for readability.

\paragraph{Expected current rewards}
Given a generated prefix, the current reward $r_t(i)$ is available for every catalog item. We therefore compute its expectation exactly under $\pi_t$:
\begin{equation}
 m_t=\sum_{i\in\mathcal I}\pi_t(i)r_t(i).
 \label{eq:revised-expectation}
\end{equation}
Optimizing $m_t$ trains the candidate distribution using both position accuracy and collaborative transition rewards. Its gradient incorporates rewards from the full candidate set, including items that were not sampled.

\paragraph{Credit from later predictions}
Selecting $x_t$ changes the context for subsequent predictions. We estimate this role using the rewards of the later prediction distributions reached after that choice. The continuation of the sampled trajectory provides the estimate $\sum_{s=t+1}^{K}m_s$. This quantity reflects both agreement with later observed items and the transitions along the generated continuation.
We combine expected current reward and continuation credit in the following surrogate loss
\begin{equation}
 {\mathcal L}_{\mathrm{rel}}
 ={}-\sum_{t=1}^{K}m_t
 -\sum_{t=1}^{K-1}
 \operatorname{sg}\!\left(
 \sum_{s=t+1}^{K}m_s-B
 \right)\log\pi_t(x_t),
 \label{eq:revised-surrogate}
\end{equation}
where $\operatorname{sg}$ denotes stop-gradient. For each training pair, we independently sample $G\geq2$ trajectories. For a given trajectory, $B$ is the mean total expected reward $\sum_s m_s$ of the other trajectories. It is independent of that trajectory's sampled choices, conditional on the training pair, and is detached in the surrogate. The loss is averaged over trajectories and training pairs.
With fixed reward functions and fresh on-policy trajectories, the negative gradient of $\mathcal{L}_{\mathrm{rel}}$ is an unbiased estimator of $\nabla_\theta J(\theta)$, provided that replay preserves all deterministic parameter dependencies.

The first term optimizes expected current rewards across all steps, with gradients propagating through probability feedback and state transitions. The second assigns continuation credit to sampled item choices through their log-probabilities. The continuation reward sum excludes the current step, whose reward is already averaged over candidate items in $m_t$. For $K=3$, $m_2$ directly trains the second-step distribution, while $m_3$ contributes to the continuation credit for $x_2$. The final step is trained through $m_K$ and has no continuation-credit term.

\paragraph{KL regularization}
We regularize the item distributions at sampled contexts toward the frozen reference $P_0$. Let $\pi_t^0$ be its item distribution conditioned on the same history and sampled prefix as $\pi_t$. Each model computes its own states and probability feedback along that shared item sequence. The Stage II loss for one sampled trajectory is
\begin{equation}
 {\mathcal L}_{\mathrm{II}}
 ={}{\mathcal L}_{\mathrm{rel}}
 +\frac{\lambda_{\mathrm{KL}}}{K}
 \sum_{t=1}^{K}\sum_{i\in\mathcal I}
 \pi_t(i)\log\frac{\pi_t(i)}{\pi_t^0(i)},
 \label{eq:revised-stage-two}
\end{equation}
where $\lambda_{\mathrm{KL}}$ controls the regularization strength. The reference probabilities are treated as constants. For this KL term, sampled items are fixed, while gradients through the current model's probabilities, feedback, and state updates are retained. Stage II uses the relation-aware surrogate and KL regularization, and the observed future enters through the rewards.

\paragraph{Optimization}

At each update, we sample fresh trajectories and replay the sampled item sequences with gradients enabled, using the same model parameters for sampling and replay. Replay recomputes probabilities, probability feedback, and behavioral states, preserving their deterministic parameter dependencies in both $m_t$ and $\log\pi_t(x_t)$.
We average ${\mathcal L}_{\mathrm{II}}$ over trajectories and training pairs and take one optimizer step. The encoder, SID heads, and transition head are updated jointly. SID assignments, the feedback table, reward functions, and $P_0$ remain fixed.

\subsection{Greedy Inference and Complexity}
\label{subsec:revised-inference}

\paragraph{Greedy inference}
The history is encoded once to obtain $h_1$. At each future step, we compute the catalog distribution and select 
\begin{equation}
 x_t=\arg\max_{i\in\mathcal I}\pi_t(i).
 \label{eq:revised-greedy}
\end{equation}
For $t<K$, the selected item and the full distribution $\pi_t$ produce the feedback and the next behavioral state. The output is the ordered sequence $(x_1,\ldots,x_K)$. 

\paragraph{Time complexity}
Let $V$ be the vocabulary size of each SID codebook and $d$ the model width.
We assume fixed-depth dense residual heads and a transition MLP with
widths of order $d$, and feedback vectors of dimension $d$.
Let $C_{\mathrm{e}}(n)$ denote the cost of encoding the history,
including its input representations.
SID indices and fixed feedback vectors are stored before inference.

At each step, the $L$ residual SID heads cost $O(Ld^2)$,
token scoring costs $O(LVd)$, and item-score composition costs $O(NL)$.
Item normalization and greedy selection cost $O(N)$, which is
included in the item-scoring bound.
Each non-final step additionally requires $O(Nd)$ for probability
feedback and $O(d^2)$ for the state transition.
The total inference cost is
\begin{equation}
\small{
 O\Big(
 C_{\mathrm{e}}(n)
 +K\big[Ld^2+LVd+NL\big]
 +(K-1)\big[Nd+d^2\big]
 \Big).}
 \label{eq:revised-complexity}
\end{equation}
For fixed SID settings, the catalog-dependent computation is linear in $N$, and generation is linear in $K$.

\paragraph{Memory and training cost}
The stored SID indices and feedback table occupy $O(NL+Nd)$ memory,
in addition to model parameters and the encoder workspace.
Greedy decoding reuses per-step buffers and does not require storing
all $K$ catalog distributions.
Stage II adds sampling, differentiable replay, and a complete frozen
reference forward pass for each trajectory.
The reference pass has the same asymptotic dependence on $N$ and $K$
as generation.
Given the two item distributions, the KL sum itself costs $O(N)$
per step.
With the reward scores available, computing $m_t$ also costs
$O(N)$, and all continuation sums are obtained in $O(K)$ time per
trajectory by a reverse cumulative sum.
The costs of constructing and storing the collaborative reward statistics are
separate offline costs.

\section{Experiments}

We evaluate \modelname{} on ordered future interaction prediction, focusing on prediction quality, the contributions of its main components, and computational cost. Our experiments address the following five research questions:
\begin{description}
    \item[RQ1.]
    How does our model compare with existing methods in predicting user behavior trajectories?

    \item[RQ2.]
    How do behavioral state updates and ordered supervision affect trajectory prediction?

    \item[RQ3.]
    How do different probability feedback rules affect trajectory prediction?

    \item[RQ4.]
    How do collaborative transition rewards and continuation credit contribute to trajectory prediction?

    \item[RQ5.]
    What are the inference time and peak memory costs of \modelname{} relative to other baselines?
\end{description}

\subsection{Experimental Setup}
\label{subsec:exp-setup}
\paragraph{Data and evaluation}
We evaluate \modelname{} on Beauty, Cell Phones \& Accessories, ML-1M, and Last-FM. Table~\ref{tab:dataset_statistics} lists the statistics of the four datasets used in our experiments. RUsers denotes the retained cohort used during training and evaluation in our local experiments at the specified $K$. Following the previous standard~\cite{huang2025listwise,huang2026generative},  we use $K=3$ for Beauty and Cell Phones \& Accessories, which have shorter average user sequences, and $K=5$ for ML-1M and Last-FM, which have longer sequences. These common settings facilitate fair comparisons at matched prediction horizons. We adopt a chronological trajectory leave-out evaluation setup, following LPDO and GCB. Each example consists of an observed interaction history and the next $K$ interactions in their original temporal order. For each dataset and prediction horizon, all locally evaluated methods use the same preprocessed interaction sequences, data splits, and history--future windows. Training windows and collaborative transition statistics are constructed exclusively from the training portion. For a chronological user sequence $S_u=(i_{u,1},\ldots,i_{u,T_u})$, the final $K$ interactions form the test target, and the immediately preceding $K$ interactions form the validation target. Training pairs are constructed only from $S_u[1:T_u-2K]$. The validation and test histories are $S_u[1:T_u-2K]$ and $S_u[1:T_u-K]$, respectively.
We report MHR@M, MNDCG@M, SHR@M, and SNDCG@M with $M\in\{5,10\}$, and SeqMatch@50 to measure the fraction of test trajectories for which every target item appears among the top-50 candidates at its corresponding future position. For each future position, HR and NDCG are computed over the test instances. MHR and MNDCG take the arithmetic mean of these position-wise scores, whereas SHR and SNDCG take their geometric mean.

Thus, $K$ denotes the trajectory length, while $M$ denotes the ranking cutoff at each future position. All locally evaluated methods use the same eligible item catalog, without sampled negatives. Models that score the full catalog are evaluated by their catalog ranks, whereas beam-based generators are evaluated using their decoded candidate lists, as described below.
For beam-based SID generators, including TIGER and our GCB implementation, we follow GCB's trajectory decoding and evaluation route, using beam width 100. Generated token sequences are divided into consecutive SID blocks and mapped back to catalog items through the item-code dictionary. Beams containing invalid item mappings are discarded before evaluation. Position-wise ranking metrics are computed from the decoded item candidates. SeqMatch@50 is computed from the same position-wise ranked lists using the definition above.
\begin{table}[t]
  \centering
  \caption{Statistics of the datasets used in our experiments.}
  \label{tab:dataset_statistics}
\resizebox{1\linewidth}{!}{%
  \begin{tabular}{lcccccc}
  \hline
  Dataset & Users & RUsers & Items & Interactions & AvgLen & $K$ \\
  \hline
  Beauty          & 22,363 &  5,123 & 12,101 & 198,502   & 8.88    & 3 \\
  Cell Phones     & 27,879  &  3,216 & 10,429 & 194,439    & 6.97   & 3 \\
  MovieLens-1M    & 6,040 & 6,040 & 3,706  & 1,000,209   & 165.60  & 5 \\
  Last-FM          & 891   & 829 & 987  & 1,048,575 & 1176.85 & 5 \\
  \hline
  \end{tabular}}
\end{table}

\paragraph{Evaluation metrics}
Let $N_{\mathrm{test}}$ be the number of test trajectories, $K$ the prediction horizon, and $M$ the ranking cutoff. For test trajectory $n$ at future position $t$, let $r_{n,t}$ denote the one-based rank of the ground-truth item. Since each position has one relevant item, its hit and discounted ranking scores are
\begin{equation}
\begin{aligned}
H_{n,t} = \mathbf{1}\{r_{n,t}\leq M\},\quad
D_{n,t} = \frac{H_{n,t}}{\log_2(r_{n,t}+1)}.
\end{aligned}
\end{equation}
We first compute the position-wise test scores:
\begin{equation}
\begin{aligned}
h_t = \frac{1}{N_{\mathrm{test}}}
       \sum_{n=1}^{N_{\mathrm{test}}} H_{n,t}, \quad
d_t = \frac{1}{N_{\mathrm{test}}}
       \sum_{n=1}^{N_{\mathrm{test}}} D_{n,t}.
\end{aligned}
\end{equation}
MHR and MNDCG use arithmetic aggregation across future positions, while SHR and SNDCG use geometric aggregation~\citep{huang2025listwise}:
\begin{equation}
\begin{aligned}
\mathrm{MHR}@M
  &= \frac{1}{K}\sum_{t=1}^{K} h_t, \quad
\mathrm{MNDCG}@M
  = \frac{1}{K}\sum_{t=1}^{K} d_t,\\
\mathrm{SHR}@M
  &= \left(\prod_{t=1}^{K}h_t\right)^{\frac{1}{K}},\quad
\mathrm{SNDCG}@M
  = \left(\prod_{t=1}^{K}d_t\right)^{\frac{1}{K}}.
\end{aligned}
\end{equation}
We additionally report SeqMatch, the fraction of test trajectories for which every target item appears among the top-$M$ candidates at its corresponding position:
\begin{equation}
\mathrm{SeqMatch}@M
=
\frac{1}{N_{\mathrm{test}}}
\sum_{n=1}^{N_{\mathrm{test}}}
\prod_{t=1}^{K}H_{n,t}.
\end{equation}
All metrics are higher-is-better.

\paragraph{Baselines}
We compare \modelname{} with 11 baseline methods and two additional variants, SASRec-M and RPG-M, covering conventional sequential recommendation, diffusion-based recommendation, generative recommendation, and user behavior trajectory prediction. FPMC~\cite{FPMC} combines matrix factorization with first-order Markov transitions.   SASRec~\cite{kang2018self} uses self-attention to model interaction histories, while STOSA~\cite{fan2022sequential} incorporates stochastic representations   of item transitions.  DCRec~\cite{yang2023debiased} augments sequential representations with contrastive learning. Among diffusion-based methods,   DiffuRec~\cite{li2023diffurec} reconstructs target-item representations through history-conditioned reverse diffusion, and   PreferDiff~\cite{liu2025preference} incorporates personalized ranking signals into diffusion training. LPDO~\cite{huang2025listwise} formulates UBTP as diffusion-based trajectory generation with listwise preference supervision. For generative recommendation, TIGER~\cite{rajput2023recommender}   represents items with residual-quantized semantic IDs and autoregressively generates the next item's SID. RPG~\cite{hou2025generating} uses   unordered semantic IDs to support parallel token prediction within an item. EchoRec~\cite{ma2026echorec} uses future behaviors as supervision and aligns   holistic preference representations with the decoding representations. GCB~\cite{huang2026generative} represents items with discrete semantic IDs and   autoregressively generates a sequence of future behaviors.

FPMC, STOSA, DCRec, DiffuRec, PreferDiff, SASRec, RPG, TIGER, and EchoRec retain their model-specific training objectives and are adapted to the $K$-step UBTP task. FPMC, SASRec, STOSA, DCRec, and DiffuRec produce position-wise rankings directly from the observed history, without feeding generated items back. PreferDiff, RPG, and EchoRec use greedy autoregressive rollout, appending each top-1 generated item before the next step, whereas TIGER uses autoregressive beam search and extends multiple candidate paths. RPG, TIGER, and EchoRec retain their respective within-item SID prediction and decoding procedures. Separately, SASRec-M and RPG-M are trained with multi-step objectives supervised at all $K$ future positions and are evaluated autoregressively. We compute each model's ranking metrics from its position-wise outputs, and TIGER's ranks are based on its decoded 100 beam candidates.  Generated SID blocks are mapped to catalog items, and beams containing invalid item mappings are discarded. At each future position, repeated occurrences of the same item across valid beams are merged. If fewer than $M$ distinct valid items remain, the available list is evaluated without padding or additional search. Ground-truth future items are never used as validation or test inputs. We reproduce LPDO using the authors' publicly available implementation\footnote{\url{https://github.com/Fr1ck/LPDO}} and recompute its ranking metrics with the common evaluator used for the other methods. The reported LPDO results are obtained from this reproduction under our unified evaluation protocol. 
No public implementation of GCB is available. 
So we implement GCB from the paper's description, using RQ-VAE semantic IDs and an autoregressive Transformer, and decode with beam size 100.  For all locally evaluated methods, we use the same registered data splits, rank items over the same full catalog. Checkpoint selection uses validation SNDCG@10, and the test split is not used for checkpoint or configuration selection. We follow author-released configurations or documented settings where available and make validation-based choices otherwise.

\paragraph{Implementation Details}
We implement our model in PyTorch and conduct our experiments on one NVIDIA H200 GPU, using Adam for optimization. Training consists of Stage I supervised initialization followed by Stage II relation-aware optimization. Table~\ref{tab:implementation} summarizes the model configuration, the optimization settings for each stage, and the coefficients used for feedback and reward-based learning. \modelname{} uses greedy decoding at inference.  For the methods and ablation variants reproduced by us, we report mean performance over three training runs with random seeds 42, 123, and 2024.  Furthermore, the reported improvements of the main metric are statistically significant (p $<$ 0.05).

  \begin{table}[t]
  \centering
  \caption{Implementation Details}
  \label{tab:implementation}
  \footnotesize
  \setlength{\tabcolsep}{3pt}
  \renewcommand{\arraystretch}{1.08}
  \begin{tabularx}{\columnwidth}{@{}
    >{\raggedright\arraybackslash}p{0.39\columnwidth}
    >{\raggedright\arraybackslash}X@{}}
  \toprule
  Parameter & Configuration \\
  \midrule
\multicolumn{2}{@{}l}{\textbf{Architecture}} \\
  Maximum history length & 50 items \\
  SID tokens / codebook size & 16 tokens per item /  256 codes per position \\
  Item / feedback  dimensions & 448 / 448  \\
  Encoder & 2-layer causal Transformer, 4 heads, FFN width 1024  \\
  \midrule
  \multicolumn{2}{@{}l}{\textbf{Stage I}} \\
  Learning rate & $10^{-4}$ \\
  Batch size & 256 \\
  Maximum epochs & 80 \\
   Temperature $\tau_{\mathrm{sid}}$ / $\tau_{\mathrm{item}}$ & 0.03 /  0.15 \\  %
   Loss weight \(\lambda_{\mathrm{item}}\)/ \(\lambda_{\mathrm{traj}}\)&1 / 1\\
  \midrule
  \multicolumn{2}{@{}l}{\textbf{Stage II}} \\
  Learning rate & $10^{-5}$ \\
  Batch size & 32 \\
  Maximum epochs  & 10\\
  Rollouts per history & 8 \\
  \midrule
  \multicolumn{2}{@{}l}{\textbf{Probability feedback and regularization}} \\
  Mixing coefficient $\eta$ & Stage I: 0 /  Stage II and evaluation: 0.75 \\
  Similarity temperature & $\tau_{\mathrm{sim}}=0.2$ \\
  Transition reward weight & $\lambda_A=0.25$ \\
  KL coefficient & $\lambda_{KL}=0.5$ \\
  \bottomrule
  \end{tabularx}
  \end{table}

\subsection{Overall Trajectory Prediction (RQ1)}
\label{subsec:exp-overall}
\begin{table*}[t]
\centering
\caption{Overall UBTP results on four datasets. MHR, MNDCG, SHR, and SNDCG are evaluated at cutoffs 5 and 10, and SeqMatch is evaluated at cutoff 50. Bold values indicate the highest reported score for each dataset and metric.}
\label{tab:mainresult}
\resizebox{1\textwidth}{!}{%
\begin{tabular}{lcccccccccccccc}
\hline
Metric & FPMC & SASRec & SASRec-M & STOSA & DiffuRec & PreferDiff & DCRec & TIGER & RPG & RPG-M & LPDO & EchoRec & GCB & Ours \\
\hline
\multicolumn{15}{l}{\textbf{Beauty ($K=3$)}} \\
MHR@5 & 0.0195 & 0.0182 & 0.0124 & 0.0156 & 0.0183 & 0.0262 & 0.0020 & 0.0035 & 0.0250 & 0.0353 & 0.0285 & 0.0312 & 0.0192 & \textbf{0.0381} \\
MNDCG@5 & 0.0123 & 0.0113 & 0.0074 & 0.0103 & 0.0115 & 0.0187 & 0.0011 & 0.0023 & 0.0151 & 0.0228 & 0.0186 & 0.0213 & 0.0121 & \textbf{0.0252} \\
MHR@10 & 0.0324 & 0.0289 & 0.0215 & 0.0221 & 0.0333 & 0.0346 & 0.0042 & 0.0059 & 0.0427 & 0.0539 & 0.0450 & 0.0472 & 0.0268 & \textbf{0.0629} \\
MNDCG@10 & 0.0164 & 0.0148 & 0.0104 & 0.0125 & 0.0163 & 0.0214 & 0.0018 & 0.0031 & 0.0207 & 0.0288 & 0.0239 & 0.0265 & 0.0146 & \textbf{0.0332} \\
SHR@5 & 0.0191 & 0.0174 & 0.0118 & 0.0152 & 0.0172 & 0.0196 & 0.0018 & 0.0034 & 0.0217 & 0.0307 & 0.0270 & 0.0269 & 0.0185 & \textbf{0.0335} \\
SNDCG@5 & 0.0120 & 0.0108 & 0.0071 & 0.0100 & 0.0106 & 0.0131 & 0.0011 & 0.0023 & 0.0128 & 0.0193 & 0.0176 & 0.0180 & 0.0115 & \textbf{0.0217} \\
SHR@10 & 0.0319 & 0.0279 & 0.0213 & 0.0214 & 0.0320 & 0.0273 & 0.0042 & 0.0057 & 0.0390 & 0.0482 & 0.0425 & 0.0423 & 0.0260 & \textbf{0.0577} \\
SNDCG@10 & 0.0161 & 0.0142 & 0.0102 & 0.0121 & 0.0154 & 0.0156 & 0.0018 & 0.0030 & 0.0184 & 0.0251 & 0.0226 & 0.0230 & 0.0140 & \textbf{0.0296} \\
SeqMatch@50 & 0.0244 & 0.0228 & 0.0094 & 0.0064 & 0.0240 & 0.0025 & 0.0000 & 0.0014 & 0.0217 & 0.0336 & 0.0322 & 0.0230 & 0.0039 & \textbf{0.0435} \\
\hline
\multicolumn{15}{l}{\textbf{Cell Phones \& Accessories ($K=3$)}} \\
MHR@5 & 0.0109 & 0.0044 & 0.0033 & 0.0092 & 0.0161 & 0.0155 & 0.0083 & 0.0062 & 0.0169 & 0.0170 & 0.0153 & 0.0235 & 0.0108 & \textbf{0.0281} \\
MNDCG@5 & 0.0068 & 0.0027 & 0.0023 & 0.0064 & 0.0108 & 0.0107 & 0.0056 & 0.0038 & 0.0110 & 0.0111 & 0.0096 & 0.0156 & 0.0073 & \textbf{0.0179} \\
MHR@10 & 0.0196 & 0.0074 & 0.0062 & 0.0153 & 0.0239 & 0.0219 & 0.0150 & 0.0091 & 0.0261 & 0.0274 & 0.0257 & 0.0377 & 0.0146 & \textbf{0.0483} \\
MNDCG@10 & 0.0096 & 0.0036 & 0.0033 & 0.0084 & 0.0133 & 0.0128 & 0.0077 & 0.0047 & 0.0140 & 0.0144 & 0.0129 & 0.0201 & 0.0086 & \textbf{0.0244} \\
SHR@5 & 0.0108 & 0.0043 & 0.0031 & 0.0089 & 0.0130 & 0.0121 & 0.0078 & 0.0051 & 0.0158 & 0.0158 & 0.0151 & 0.0224 & 0.0106 & \textbf{0.0264} \\
SNDCG@5 & 0.0068 & 0.0026 & 0.0022 & 0.0062 & 0.0082 & 0.0083 & 0.0053 & 0.0033 & 0.0101 & 0.0101 & 0.0094 & 0.0149 & 0.0072 & \textbf{0.0168} \\
SHR@10 & 0.0190 & 0.0072 & 0.0059 & 0.0149 & 0.0210 & 0.0181 & 0.0142 & 0.0073 & 0.0247 & 0.0261 & 0.0252 & 0.0362 & 0.0141 & \textbf{0.0454} \\
SNDCG@10 & 0.0094 & 0.0036 & 0.0031 & 0.0081 & 0.0109 & 0.0102 & 0.0074 & 0.0040 & 0.0130 & 0.0134 & 0.0126 & 0.0193 & 0.0084 & \textbf{0.0229} \\
SeqMatch@50 & 0.0050 & 0.0003 & 0.0006 & 0.0003 & 0.0065 & 0.0012 & 0.0009 & 0.0000 & 0.0062 & 0.0065 & 0.0078 & 0.0115 & 0.0022 & \textbf{0.0202} \\
\hline
\multicolumn{15}{l}{\textbf{ML-1M ($K=5$)}} \\
MHR@5 & 0.0193 & 0.0852 & 0.0704 & 0.0558 & 0.0888 & 0.0387 & 0.0794 & 0.0821 & 0.0600 & 0.1108 & 0.1142 & 0.0599 & 0.0885 & \textbf{0.1363} \\
MNDCG@5 & 0.0111 & 0.0527 & 0.0440 & 0.0340 & 0.0553 & 0.0248 & 0.0477 & 0.0525 & 0.0380 & 0.0721 & 0.0720 & 0.0381 & 0.0539 & \textbf{0.0870} \\
MHR@10 & 0.0418 & 0.1473 & 0.1183 & 0.0980 & 0.1488 & 0.0604 & 0.1437 & 0.1350 & 0.0988 & 0.1745 & 0.1871 & 0.0977 & 0.1271 & \textbf{0.2144} \\
MNDCG@10 & 0.0182 & 0.0727 & 0.0593 & 0.0475 & 0.0746 & 0.0318 & 0.0683 & 0.0694 & 0.0504 & 0.0926 & 0.0954 & 0.0503 & 0.0665 & \textbf{0.1120} \\
SHR@5 & 0.0192 & 0.0849 & 0.0687 & 0.0557 & 0.0886 & 0.0286 & 0.0763 & 0.0778 & 0.0538 & 0.1020 & 0.1121 & 0.0531 & 0.0845 & \textbf{0.1300} \\
SNDCG@5 & 0.0110 & 0.0526 & 0.0428 & 0.0339 & 0.0551 & 0.0175 & 0.0458 & 0.0490 & 0.0332 & 0.0651 & 0.0705 & 0.0332 & 0.0511 & \textbf{0.0824} \\
SHR@10 & 0.0417 & 0.1467 & 0.1157 & 0.0979 & 0.1482 & 0.0478 & 0.1391 & 0.1292 & 0.0918 & 0.1647 & 0.1839 & 0.0899 & 0.1250 & \textbf{0.2076} \\
SNDCG@10 & 0.0182 & 0.0724 & 0.0579 & 0.0475 & 0.0743 & 0.0237 & 0.0659 & 0.0656 & 0.0455 & 0.0854 & 0.0936 & 0.0451 & 0.0645 & \textbf{0.1074} \\
SeqMatch@50 & 0.0162 & 0.1103 & 0.0685 & 0.0586 & 0.1098 & 0.0020 & 0.0884 & 0.0442 & 0.0397 & 0.0949 & 0.1449 & 0.0353 & 0.0068 & \textbf{0.1545} \\
\hline
\multicolumn{15}{l}{\textbf{Last-FM ($K=5$)}} \\
MHR@5 & 0.1619 & 0.1966 & 0.1870 & 0.1242 & 0.1879 & 0.2200 & 0.1925 & 0.2461 & 0.0082 & 0.2265 & 0.2159 & 0.0077 & 0.2094 & \textbf{0.2531} \\
MNDCG@5 & 0.1133 & 0.1404 & 0.1304 & 0.0928 & 0.1370 & 0.1705 & 0.1366 & 0.1827 & 0.0042 & 0.1762 & 0.1586 & 0.0048 & 0.1538 & \textbf{0.1855} \\
MHR@10 & 0.2343 & 0.2613 & 0.2598 & 0.1628 & 0.2610 & 0.2560 & 0.2704 & 0.3271 & 0.0154 & 0.2931 & 0.2801 & 0.0154 & 0.2533 & \textbf{0.3320} \\
MNDCG@10 & 0.1365 & 0.1612 & 0.1538 & 0.1052 & 0.1606 & 0.1822 & 0.1620 & 0.2087 & 0.0065 & 0.1978 & 0.1793 & 0.0072 & 0.1679 & \textbf{0.2111} \\
SHR@5 & 0.1614 & 0.1961 & 0.1858 & 0.1235 & 0.1871 & 0.2083 & 0.1868 & 0.2372 & 0.0071 & 0.2151 & 0.2124 & 0.0073 & 0.2050 & \textbf{0.2480} \\
SNDCG@5 & 0.1131 & 0.1399 & 0.1297 & 0.0923 & 0.1362 & 0.1598 & 0.1329 & 0.1751 & 0.0035 & 0.1662 & 0.1558 & 0.0044 & 0.1502 & \textbf{0.1795} \\
SHR@10 & 0.2339 & 0.2612 & 0.2571 & 0.1621 & 0.2601 & 0.2440 & 0.2645 & 0.3199 & 0.0145 & 0.2823 & 0.2765 & 0.0150 & 0.2490 & \textbf{0.3264} \\
SNDCG@10 & 0.1364 & 0.1610 & 0.1528 & 0.1047 & 0.1598 & 0.1715 & 0.1582 & 0.2019 & 0.0060 & 0.1881 & 0.1765 & 0.0069 & 0.1643 & \textbf{0.2052} \\
SeqMatch@50 & 0.1713 & 0.1713 & 0.0724 & 0.0712 & 0.1858 & 0.0905 & 0.1556 & 0.2014 & 0.0024 & 0.1701 & 0.1544 & 0.0036 & 0.0832 & \textbf{0.2256} \\
\hline
\end{tabular}%
}

\end{table*}

Table~\ref{tab:mainresult} compares \modelname{} with 13 baseline configurations on four datasets. \modelname{} achieves the highest reported score in all 36 dataset--metric combinations, including SeqMatch@50 on every dataset. The improvements cover both the three-step prediction tasks on Beauty and Cell Phones \& Accessories and the five-step tasks on ML-1M and Last-FM.

The strongest baseline for MNDCG@10 and SNDCG@10 varies across datasets. Relative to the strongest baseline for each dataset and metric, \modelname{} improves MNDCG@10 by approximately 1.15\%--21.39\% and SNDCG@10 by approximately 1.63\%--18.65\%. Its advantage therefore holds against different leading competitors across the four datasets. The gains also extend across both ranking cutoffs and both arithmetic aggregation (MHR and MNDCG) and geometric aggregation (SHR and SNDCG), showing that the comparative advantage is retained under different summaries of position-wise performance.

On SeqMatch@50, the relative improvements are approximately 29.46\%, 75.65\%, 6.63\%, and 12.02\% on Beauty, Cell Phones \& Accessories, ML-1M, and Last-FM, respectively. These gains indicate that more test trajectories have every target item included in the top-50 candidate list at its corresponding future position. This complements the position-aggregated ranking results. On Last-FM, for example, the relatively modest improvements in MNDCG@10 and SNDCG@10 are accompanied by a larger relative gain in SeqMatch@50, demonstrating the additional benefit observed when coverage is assessed jointly across all target positions.

\begin{figure}[t]
  \centering
  \includegraphics[width=1\columnwidth]{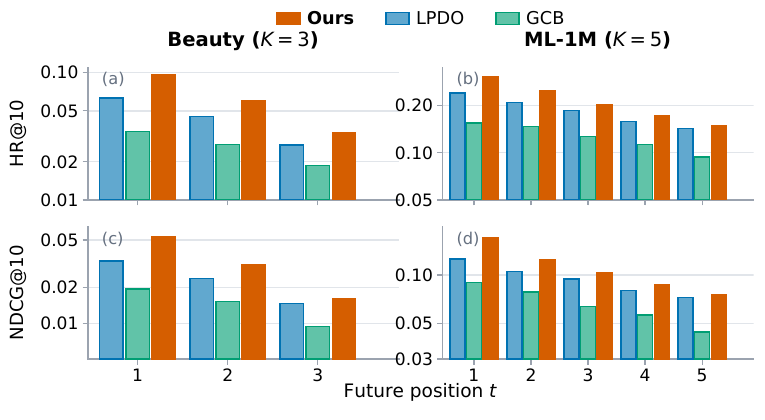}
  \caption{Comparison of three representative methods at each future trajectory position on Beauty and ML-1M, measured by HR@10 and NDCG@10.}
  \label{fig:positionwise}
\end{figure}

\textbf{Position-wise prediction.} We select LPDO and GCB as representative diffusion-based and autoregressive SID-based methods for trajectory prediction, respectively.  As shown in Fig.~\ref{fig:positionwise}, \modelname{} achieves the highest HR@10 and NDCG@10 at every position on both datasets. 

\subsection{Behavioral State Evolution (RQ2)}
\label{subsec:exp-state}

\begin{table*}[t]
  \caption{State evolution and feedback comparisons on Beauty and ML-1M. The best result in each column  within each panel is shown in bold.}
  \label{tab:exp-generation}
  \centering
  \small
  \begin{tabular*}{1\textwidth}{
    @{\extracolsep{\fill}}lcccccccc@{}
  }
    \toprule
    & \multicolumn{4}{c}{Beauty ($K=3$)}
    & \multicolumn{4}{c}{ML-1M ($K=5$)} \\
    \cmidrule(lr){2-5}
    \cmidrule(lr){6-9}
    Variant
    & \shortstack{SNDCG@5}
    & \shortstack{SHR@5}
    & \shortstack{SNDCG@10}
    & \shortstack{SHR@10}
    & \shortstack{SNDCG@5}
    & \shortstack{SHR@5}
    & \shortstack{SNDCG@10}
    & \shortstack{SHR@10} \\
    \midrule

    \multicolumn{9}{@{}l}{
      \textit{A. State evolution and supervision (RQ2)}
    } \\

    History Only
    & 0.0165 & 0.0263 & 0.0219 & 0.0429
    & 0.0384 & 0.0612 & 0.0521 & 0.1034 \\

    Zero Feedback
    & 0.0165 & 0.0259 & 0.0219 & 0.0428
    & 0.0395 & 0.0633 & 0.0528 & 0.1048 \\

    Multi-label
    & 0.0115 & 0.0194 & 0.0163 & 0.0343
    & 0.0288 & 0.0461 & 0.0401 & 0.0812 \\

    Stage I Only
    & \textbf{0.0201} & \textbf{0.0328}
    & \textbf{0.0258} & \textbf{0.0501}
    & \textbf{0.0763} & \textbf{0.1108}
    & \textbf{0.0972} & \textbf{0.1823} \\

    \midrule

    \multicolumn{9}{@{}l}{
      \textit{B. Probability feedback design (RQ3)}
    } \\

    Item Only
    & 0.0202 & 0.0320 & 0.0273 & 0.0510
    & 0.0779 & 0.1209 & 0.0972 & 0.1851 \\

    Item + Prob
    & 0.0207 & 0.0324 & 0.0281 & 0.0536
    & 0.0820 & 0.1269 & 0.1051 & 0.2045 \\

    Item + Sim
    & 0.0201 & 0.0302 & 0.0262 & 0.0490
    & 0.0738 & 0.1184 & 0.0978 & 0.1927 \\
    
    w/o Item Anchor			 
    & 0.0214 & 0.0331 & 0.0292 & 0.0569
    & 0.0809 & 0.1285 & 0.1059	& 0.2047 \\
    
    Full
    & \textbf{0.0217} & \textbf{0.0335}
    & \textbf{0.0296} & \textbf{0.0577}
    & \textbf{0.0824} & \textbf{0.1300}
    & \textbf{0.1074} & \textbf{0.2076} \\

    \bottomrule
  \end{tabular*}
\end{table*}

Panel A of Table~\ref{tab:exp-generation} compares four supervised variants in Stage I. \emph{History Only} predicts each future position from the same history representation and its position embedding, without recurrent state updates. \emph{Zero Feedback} retains the state recurrence but sets the item feedback input to zero. \emph{Multi-label} uses the History Only architecture and replaces the ordered trajectory objective with multi-label BCE over the future item set, while retaining the auxiliary next item loss.  \emph{Stage I Only} evaluates the checkpoint before Stage-II fine-tuning using the same probability feedback generation procedure and feedback coefficient as the full model. 
Zero Feedback produces almost no improvement on Beauty and only small gains on ML-1M over History Only. Compared with History Only, Stage I Only increases SNDCG@10 from 0.0219 to 0.0258 on Beauty and from 0.0521 to 0.0972 on ML-1M.  Among the ordered variants, the main improvement therefore comes from supplying item information to the state updates.
Multi-label obtains the lowest scores on both datasets. Relative to History Only, its SNDCG@10 decreases by approximately 25\% on Beauty and 23\% on ML-1M.

\subsection{Probability Feedback (RQ3)}
\label{subsec:exp-feedback}

Panel B of Table~\ref{tab:exp-generation} compares multiple feedback rules used to update the behavioral state. Each variant is trained separately in Stage II, with its feedback rule used consistently during both training and inference. 
\emph{Item Only} uses the selected-item representation alone. Item + Probability (\emph{Item + Prob}) combines this representation with a candidate summary weighted by the current predicted probabilities, without similarity reweighting. Item + Similarity (\emph{Item + Sim}) combines the selected-item representation with a candidate summary weighted by similarity to the selected item, without using the predicted probabilities in these weights. \emph{w/o Item Anchor}  retains the probability and similarity weighting while removing the explicit selected-item contribution.  Full uses both the predicted probabilities and item similarity to determine the weights of the candidate summary. The mixing coefficient is kept identical across the three variants with an additional summary, while the remaining model and training settings are held fixed.

Item + Prob outperforms Item Only across all reported metrics on both datasets, supporting the use of a probability-weighted candidate summary in addition to the selected-item representation. Item + Sim does not provide uniform improvements: it underperforms Item Only on Beauty and produces mixed results on ML-1M. Full achieves the highest values across all reported metrics. In particular, compared with Item + Prob, Full improves SNDCG@10 and SHR@10 by 5.34\% and 7.65\% on Beauty. Since both variants retain selected-item feedback and probability weighting, this comparison supports the additional benefit of similarity reweighting. Overall, the results favor the joint weighting used by \modelname{} over either probability-based or similarity-based weighting alone. w/o Item Anchor obtains lower scores, supporting the inclusion of the selected item alongside the feedback embedding weighted by predicted probabilities and similarity.

\textbf{Prediction quality across future positions.} 
\begin{figure}
    \centering
    \includegraphics[width=1\linewidth]{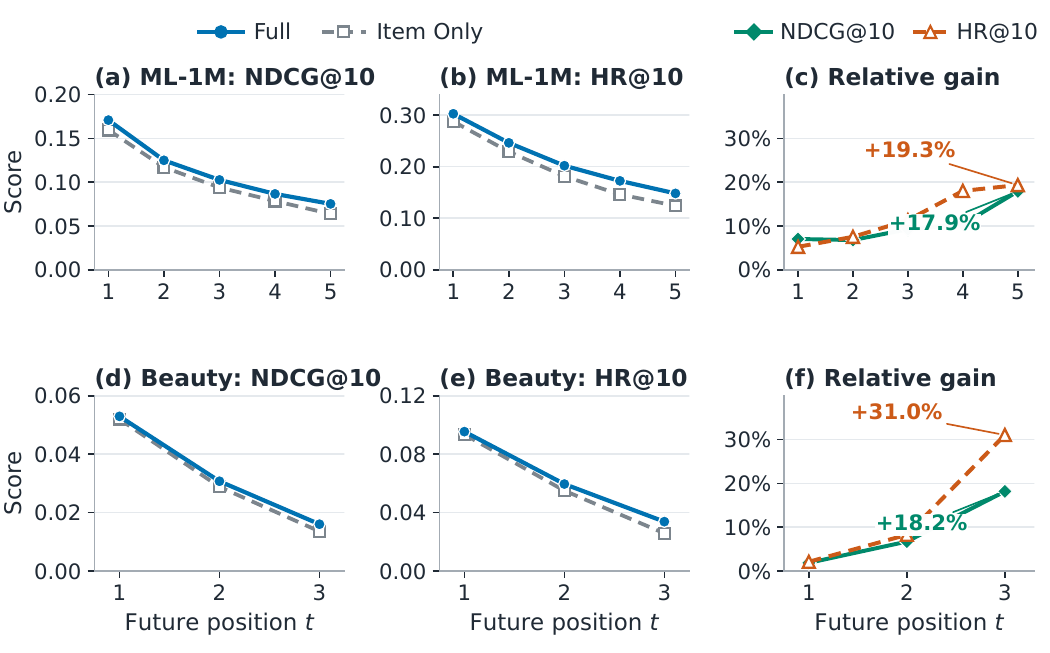}
    \caption{Prediction quality at each future position on   ML-1M  and Beauty.   The first two columns report position-wise NDCG@10 and HR@10 for Full and Item Only.  The last column shows the relative improvement of Full,   computed as  $100(\mathrm{Full}/\mathrm{ItemOnly}-1)$   for each metric and position.}
    \label{fig:exp-future-position}
\end{figure}
Fig.~\ref{fig:exp-future-position} compares position-wise NDCG@10 and HR@10 for Full and Item Only. Although both variants decline as the prediction horizon advances, Full achieves higher scores at every position on both datasets. On ML-1M, the relative HR@10 gain increases from approximately 5\% at the first position to 19.3\% at the fifth, while the NDCG@10 advantage is generally larger at later positions. On Beauty, the relative gains increase from approximately 2\% for both metrics at the first position to 18.2\% in NDCG@10 and 31.0\% in HR@10 at the third. These results suggest that probability feedback benefits prediction quality at later rollout steps.

\subsection{Relation-Aware Learning (RQ4)}
\label{subsec:exp-credit}
\begin{table*}[t]
  \caption{Ablation of relation-aware learning on Beauty and ML-1M.
  CT and CC denote the collaborative transition reward and continuation
  credit. Dashes indicate that CT and CC are not applicable to the supervised variants.
  Best scores are in bold.}
  \label{tab:exp-optimization}
  \centering
  \footnotesize
  \setlength{\tabcolsep}{2.8pt}
  \renewcommand{\arraystretch}{1.12}
  \begin{tabular*}{\textwidth}{@{\extracolsep{\fill}}lcccccccccc@{}}
    \toprule
    & & & \multicolumn{4}{c}{Beauty ($K=3$)}
        & \multicolumn{4}{c}{ML-1M ($K=5$)} \\
    \cmidrule(lr){4-7}\cmidrule(l){8-11}
    Variant & CT & CC
     & \shortstack{SNDCG@5} & \shortstack{SHR@5}
    & \shortstack{SNDCG@10} & \shortstack{SHR@10}
    & \shortstack{SNDCG@5} & \shortstack{SHR@5}
    & \shortstack{SNDCG@10} & \shortstack{SHR@10} \\
    \midrule
    Stage I Only & --- & ---
    & 0.0201 & 0.0328 & 0.0258 & 0.0501
    & 0.0763 & 0.1108 & 0.0972 & 0.1823 \\
    Supervised FT & --- & ---
    & 0.0201 & 0.0325 & 0.0261 & 0.0505
    & 0.0805 & 0.1211 & 0.1034 & 0.1863 \\
    Generated-Prefix FT & --- & ---
    &0.0204	&0.0316	&0.0265	&0.0512
    &0.0802	&0.1217	&0.1055	&0.1931 \\
    \midrule
    Task Only & $\times$ & $\times$
    & 0.0212 & 0.0329 & 0.0283 & 0.0557
    & 0.0781 & 0.1276 & 0.1012 & 0.1968 \\
    Task + CC & $\times$ & $\checkmark$
    & 0.0216 & 0.0333 & 0.0285 & 0.0562
    & 0.0806 & 0.1178 & 0.1052 & 0.1933 \\
    Task + CT & $\checkmark$ & $\times$
    & 0.0204 & 0.0329 & 0.0283 & 0.0563
    & 0.0818 & 0.1229 & 0.1057 & 0.2019 \\
    \midrule
    w/o Feedback Gradients & $\checkmark$ & $\checkmark$
    & 0.0209 & 0.0324 & 0.0281 & 0.0544
    & 0.0815 & 0.1287 & 0.1067 & 0.2053 \\
    \midrule
    \textbf{Full} & $\checkmark$ & $\checkmark$
    & \textbf{0.0217} & \textbf{0.0335} & \textbf{0.0296} & \textbf{0.0577}
    & \textbf{0.0824} & \textbf{0.1300} & \textbf{0.1074} & \textbf{0.2076} \\
    \bottomrule
  \end{tabular*}
\end{table*}

Table~\ref{tab:exp-optimization} examines relation-aware trajectory learning in Stage II through supervised controls, ablations of CT and CC, and a feedback-gradient ablation. 
Here, the collaborative transition reward (CT) is the reward component that captures collaborative support for behavior transitions, whereas continuation credit (CC) is the credit-assignment mechanism that attributes subsequent expected rewards to earlier sampled choices.

\textbf{Comparison with supervised training.} \emph{Stage I Only} evaluates the generator after joint supervised initialization, without further optimization. \emph{Supervised FT} continues supervised training on observed targets, whereas \emph{Generated-Prefix FT} minimizes cross-entropy against the observed future items using the same sampled item prefixes as its trajectory-training reference. Both fine-tuning controls use the feedback configuration of \emph{Full}.
Full achieves the best scores across all reported metrics on both datasets. Its advantage over these supervised references supports the benefit of the proposed training objective beyond additional supervised updates and exposure to generated prefixes.

\textbf{Effects of CT and CC.}
The four reward-based variants train predictions across the complete horizon using model-generated trajectories, with probability feedback, state transitions, and KL regularization retained. \emph{Task Only} sets $\lambda_A=0$ and removes the continuation-credit term from Eq.~\ref{eq:revised-surrogate}. \emph{Task + CC} restores this term, whereas \emph{Task + CT} adds the collaborative reward $\lambda_A A(x_{t-1},i)$ without continuation credit. \emph{Full} includes both components, allowing the collaborative reward to contribute to both current candidate evaluation and continuation credit. Disabling CC removes only the explicit credit assigned to discrete sampled choices, and gradients from the expected rewards still propagate through probability feedback and state updates.
The effects of adding CT or CC individually vary across datasets and metrics. However, removing either component from Full lowers every reported score, supporting their complementary roles. CT favors transitions supported by collaborative interaction patterns, while CC assigns expected rewards from subsequent predictions to earlier sampled choices. Together, they connect the relation to the preceding behavior with the consequences for later predictions.

\textbf{Effect of feedback gradients.}
\emph{w/o Feedback Gradients} retains CT, CC, and the forward feedback computation of Full, but stops gradients through the predicted probabilities used to construct the feedback embedding during Stage II. The remaining objective terms and gradient paths are unchanged. Its lower scores support the additional value of the differentiable feedback path, through which subsequent losses can update earlier predictive distributions.

\subsection{Hyperparameter Sensitivity}
\label{subsec:exp-hyperparameters}

\begin{figure}
    \centering
    \includegraphics[width=0.9\linewidth]{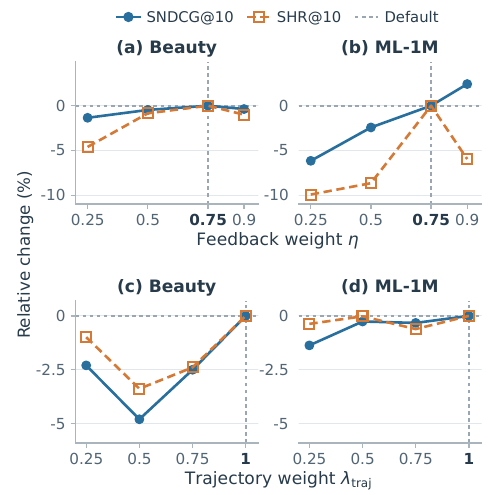}
    \caption{Sensitivity to the feedback mixing coefficient $\eta$ and trajectory-supervision weight $\lambda_{\mathrm{traj}}$. Grey lines mark the default parameter and the zero-change reference. }
    \label{fig:hypercom}
\end{figure}

We examine the feedback mixing coefficient $\eta$ and the trajectory-supervision weight $\lambda_{\mathrm{traj}}$. The figures report percentage changes relative to the corresponding default setting, allowing the sensitivity of the two metrics to be compared across Beauty and ML-1M.

\paragraph{Feedback mixing coefficient} Fig.~\ref{fig:hypercom} varies $\eta$, which controls the contribution of the soft candidate summary relative to the selected-item feedback. Increasing $\eta$ from $0.25$ to $0.75$ improves both metrics on the two datasets. Beauty obtains its strongest results at the default setting, while ML-1M exhibits a trade-off at $\eta=0.90$: SNDCG@10 increases further, but SHR@10 decreases. Thus, a larger contribution from the soft summary does not uniformly improve prediction quality. Finally, we set $\eta=0.75$.

\paragraph{Trajectory supervision weight} Fig.\ref{fig:hypercom} varies $\lambda_{\mathrm{traj}}$, which weights the trajectory loss relative to next item supervision in Stage I. The default value of $1.0$ gives the strongest overall results among the tested settings, although the changes are not monotonic. ML-1M remains comparatively stable across this range, particularly in SHR@10, for which the scores at $0.5$ and $1.0$ are nearly identical. Beauty shows greater sensitivity, with both metrics dropping at intermediate weights before recovering at the default setting. These results support retaining the default trajectory weight within the evaluated range.

\subsection{Effect of Prediction Horizon}
\label{subsec:exp-horizon}

\begin{figure}[t]
  \centering
  \includegraphics[width=0.9\linewidth]{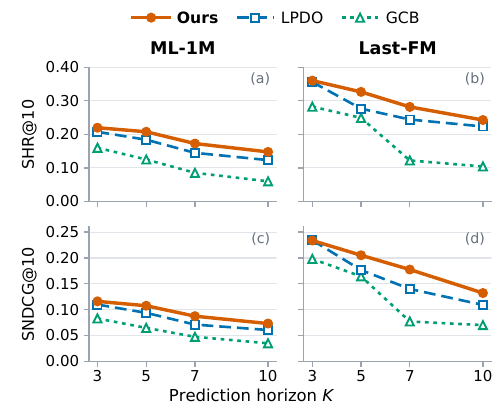}
  \caption{Comparison of three representative methods across prediction horizons $K$ on ML-1M and Last-FM. Each method is trained separately for each $K$.}
  \label{fig:exp-horizon}
\end{figure}

We examine how trajectory prediction performance changes with the prediction horizon on ML-1M and Last-FM. We compare \modelname{} with LPDO and GCB at $K\in\{3,5,7,10\}$, where $K$ denotes the number of future items to predict. Each method is trained separately for every value of $K$. For each horizon, all three methods use identical training and test samples, candidate sets, and metric implementations. Fig.~\ref{fig:exp-horizon} reports SHR@10 and SNDCG@10 for each setting, with \modelname{} labeled as \emph{Ours}. LPDO and GCB provide representative comparisons with diffusion-based and autoregressive trajectory generation, respectively. This experiment complements the main evaluation by examining whether the comparative advantage persists across separately trained models targeting different prediction horizons.
All three methods obtain lower scores as the prediction horizon increases on both datasets. \modelname{} achieves the highest SHR@10 and SNDCG@10 at every evaluated horizon on ML-1M. On Last-FM, it also achieves the highest SHR@10 throughout and the highest SNDCG@10 for $K\geq5$. At $K=3$, \modelname{} and LPDO perform closely: \modelname{} has a slight advantage in SHR@10, whereas LPDO has a marginally higher SNDCG@10. The results therefore show consistent advantages beyond the shortest horizon, while preserving a small exception in short-horizon ranking quality.

Comparing the separately trained models at $K=3$ and $K=10$, \modelname{} exhibits a smaller relative decrease in both metrics than either baseline on both datasets. Its advantage remains evident at the longest evaluated horizon, although all methods obtain substantially lower scores than at the shortest horizon. 
The size of the improvement over LPDO varies across horizons and metrics rather than increasing monotonically with $K$. Nevertheless, \modelname{} maintains the best reported scores for both metrics on both datasets at $K=5$, $7$, and $10$. Taken together, these observations support its effectiveness for longer target trajectories within the evaluated range, complementing the component-level analyses in the preceding experiments.

\subsection{Inference Efficiency (RQ5)}
\label{subsec:exp-efficiency}
\begin{figure}[t]
  \centering
  \includegraphics[width=1\linewidth]{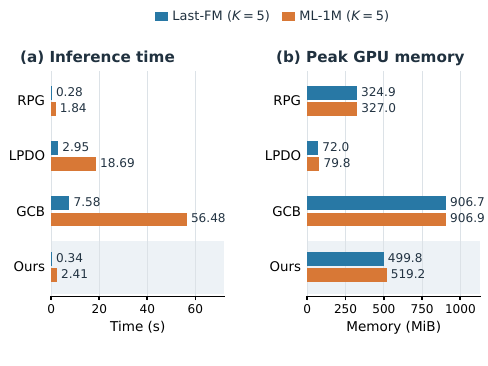}
  \caption{Comparison of inference time and peak GPU memory usage across RPG, LPDO, GCB, and TraGR on Last-FM and ML-1M.}
  \label{fig:exp-inference-efficiency}
\end{figure}

We compare \modelname{} with RPG, LPDO, and GCB on Last-FM and ML-1M. RPG provides a close architectural reference through parallel SID prediction, while LPDO and GCB represent diffusion-based and autoregressive SID-based trajectory generators, respectively. We evaluate the RPG baseline listed in Table~\ref{tab:mainresult} and use our  GCB reimplementation on both datasets. 
All measurements use a single NVIDIA H200 GPU with batch size 64. We report the total inference time over each dataset's test workload, averaged over five timed runs after three warm-up runs. The timed region includes model forward passes and trajectory generation, including LPDO denoising and GCB beam search, with GPU synchronization before and after each measurement. RPG and \modelname{} use greedy decoding, LPDO uses 1 denoising step, and GCB uses beam width 100.
As shown in Fig.~\ref{fig:exp-inference-efficiency}, \modelname{} achieves substantially lower inference time than LPDO and GCB on both datasets, with additional runtime overhead relative to RPG. Its peak GPU memory is lower than that of GCB but higher than that of LPDO. Compared with RPG, \modelname{} uses more memory on Last-FM and ML-1M. These results demonstrate a speed--memory trade-off relative to LPDO and lower inference costs in both time and memory relative to GCB under the evaluated settings.

\section{Conclusion}
We presented \modelname{}, a behavior-aware generative recommendation
framework for user behavior trajectory prediction. By coupling
behavioral state updates with item-anchored probability feedback,
the framework carries information from selected items and predictive
distributions across future positions. Relation-aware learning on
model-generated trajectories combines position accuracy and directed
collaborative transition rewards with credit assigned to earlier
choices for later prediction quality. Experiments on four public
datasets show improvements in position-wise ranking and trajectory
coverage, with ablation studies supporting the contributions of the
proposed components. Under the evaluated settings, \modelname{} also
achieves lower inference time than LPDO and GCB on ML-1M and Last-FM.
Future work will explore efficient candidate selection and feedback
aggregation to reduce catalog-dependent computation and improve
scalability to larger item catalogs.

\bibliographystyle{IEEEtran}
\bibliography{base}

\end{document}